\documentclass[preprint,onecolumn,nofootinbib]{revtex4}
\pdfoutput=1
\usepackage[colorlinks=true,linkcolor=blue,urlcolor=blue,filecolor=black,citecolor=red,pdfstartview=FitV,pdftitle={},pdfsubject={},pdfkeywords={},pdfpagemode=None,bookmarksopen=true]{hyperref}
\usepackage{graphicx}
\usepackage{amsmath}
\usepackage{amsfonts}
\usepackage{amssymb,ulem}
\usepackage{color}%
\usepackage{dcolumn}
\newcommand{\f}{\begin{equation}}
\newcommand{\ff}{\end{equation}}
\newcommand{\fa}{\begin{eqnarray}}
\newcommand{\ffa}{\end{eqnarray}}

\begin{document}
\title{Particle motion and chaos}
\author{Zhenhua Zhou $^{1}$}
\thanks{dtplanck@163.com}
\author{Jian-Pin Wu $^{2,3}$}
\thanks{jianpinwu@yzu.edu.cn}
\affiliation{$^1$~School of Physics and Electronic Information, Yunnan Normal University, Kunming, 650500, China}
\affiliation{$^2$~Center for Gravitation and Cosmology, College of Physical Science and Technology, Yangzhou University, Yangzhou 225009, China}
\affiliation{$^3$~School of Aeronautics and Astronautics, Shanghai Jiao Tong University, Shanghai 200240, China}
\begin{abstract}

In this note, we consider particle falling in the black hole with an additional potential.
Following the proposal by Susskind \cite{Susskind:2018tei},
we study the growth rate of the particle's Rindler momentum, which corresponds
to the growth rate of the operator size in the dual chaotic system. A general analysis
near the horizon shows that the growth rate of the particle's Rindler momentum of the particle falling with a regular potential is as the same as that of the particle free falling, which saturates the chaos bound.
However, when the potential is singular, the growth rate is suppressed such that it is below the Lyapunov exponent.
It implies that the chaos suppression may be captured
by an additional singular potential in the gravity side.
We further explicitly study a particle falling in hyperscaling violating spacetime to
confirm the general analysis results.
Finally we study the particle falling in AdS soliton geometry. It
also exhibits a suppression of the growth of the Rindler momentum.
It is attributed to that when the repulsive potential is introduced or the black hole horizon is absent,
the  particle is slowed down and its trajectory seen by a comoving observer is timelike,
which corresponds a weak chaos system.

\end{abstract}
\maketitle

\section{Introduction}

In \cite{Susskind:2018tei}, Susskind proposes that there is a correspondence
between the operator growth in the chaotic quantum systems
and the momentum of the particle falling toward the black hole.
In particular, they grow exponentially with the same Lyapunov exponent.
For the AdS black hole, the particle's momentum grows at a maximal rate \cite{Susskind:2018tei}
and the Lyapunov exponent saturates the chaos bound proposed in \cite{Maldacena:2015waa}.
It is a universal property because all the horizons are locally the Rindler-like.
The same characteristic is also found in the the strongly coupling chaotic quantum system, SYK model.
References \cite{Brown:2018kvn,Ageev:2018msv} further study the particle falling toward charged black holes
and confirm the Susskin proposal.

In this note, we study the growth of the particle momentum in an external potential.
In particular, we want to explore under what condition we can have a holographic dual for which the Lyapunov exponent is below the chaos bound.
Such potential can provide a platform for us to
study a realistic quantum chaos system from the gravity side.

From the near horizon analysis, it can be seen that a regular potential leads to the
same conclusion as that of a vanishing potential. In such case, the Rindler momentum exhibits a Lyapunov exponent growth.
The novel phenomena appears when the potential is singular near the horizon that the momentum growth is suppressed. According to
the duality by Susskind, it implies a chaos suppression and it is below the Lyapunov bound.
Therefore, the singular potential may capture some mechanics of the chaotic system.
Following this clue, we further explicitly study the particle falling
in hyperscaling violating (HV) black brane geometry.
By this simple example, we explicitly illuminate that the velocity bound in the gravity set the chaos bound in the quantum system.
When the free falling particle travels with the light speed near the horizon\footnote{Here, we mean that the particle nearly follows a null trajectory according the comoving observer.
It shall be explicitly illuminated in Section \ref{sec-con}.},
the Lyapunov exponent saturates the chaos bound.
If we add some repulsive potential, which results in the velocity of the particle being less than the light velocity,
the exponent of the growth of the momentum will not saturate the bound.
In this case, chaos is suppressed.

We also study the particle falling in AdS soliton geometry \cite{Witten:1998zw,Horowitz:1998ha}.
The holographic superconducting models have been built based on AdS soliton geometry \cite{Nishioka:2009zj,Horowitz:2010jq}.
Since the horizon is absent, the growth of the particle's momentum in AdS soliton geometry
is different from that in the black holes background.

\section{Particle motion and its Rindler momentum}\label{basis}

Let us consider a particle with mass $m$ and an external potential $V$ moving in $d+1$ spacetime dimensions.
It is convenient to set the metric as
\begin{align}
\label{metric}
ds^2=g_{tt}(r)dt^2+g_{rr}(r)dr^2+\ldots=-a(r)f(r)dt^2+\frac{dr^2}{b(r)f(r)}+\cdots\,.\,.
\end{align}
where $f(r)$ vanishes at the horizon $r_+$.
$a(r),b(r)$ are two positive function of $r$, which is regular at $r_+$.
The boundary locates at $r=0$.
The action of the particle is
\begin{align}
S_p=-m\int\big(1+V(X)\big)\sqrt{-g_{\mu\nu}\dot{X}^\mu \dot{X}^\nu}d\tau\,,
\label{Sp}
\end{align}
where $\tau$ is arbitrary parameter of the particle world line and the dot denotes the derivative with respect to $\tau$.
$X^{\mu}$ is the spacetime coordinates.
$V(X)$ is the external potential.
From the above action, we derive the equation of motion (EOM) as\footnote{In the non-relativistic limit, choosing $\tau=t$, we have $\dot{X}^\mu\sim 1/c,\,\mu=1,2,3$, $V\sim1/c^2$. Thus $\dot{V}\dot{X}^\mu\sim 1/c^3$
can be neglected and $\eta\approx1$, $1+V\approx1$. Therefore, the EOM \eqref{eomv1} reduces to the usual Newton law, $F=m\ddot{x}=-\nabla V$, which is expected.}
\begin{subequations}
\label{eom}
\begin{align}
&\ddot{X}^\mu+\Gamma_{\alpha\beta}^\mu\dot{X}^\alpha \dot{X}^\beta-\frac{\dot{\eta} \dot{X}^\mu}{2\eta}+\frac{\dot{V}\dot{X}^\mu+\eta g^{\mu\nu}\partial_\nu V}{1+V}=0\,,\label{eomv1}\\
&\eta\equiv-g_{\alpha\beta}\dot{X}^\alpha \dot{X}^\beta>0\,,
\end{align}
\end{subequations}
Here we only assume the timelike condition that $\eta>0$.
We would like to point out that $\eta$ is not a constant in general because $\tau$ is an arbitrary parameter of the particle world line.
Only when $\tau$ is the affine parameter, $\eta=1$.
And then, we obtain the canonical momentum
\begin{align}
p_\mu=\frac{\delta S_p}{\delta \dot{X}^\mu  }=\frac{m\big(1+V(X)\big)\,g_{\mu\nu} \dot{X}^\nu}{\sqrt{-g_{\alpha\beta}\dot{X}^\alpha \dot{X}^\beta}}\,.
\label{p}
\end{align}

To proceed, we choose the static gauge $\tau=t$ and take the ansatz $r=r(t)$ and $x,y=const.$ in what follows.
In addition, we assume that the potential only depends on $r$, i.e., $V=V(r)$.
And then, Eqs.\eqref{eom} and \eqref{p} reduce to
\begin{subequations}
\label{sol}
\begin{align}
&\dot{r}=\sqrt{ab}f(1-af\frac{(1+V)^2}{A^2})^{1/2}\,,\label{sol-r}\\
&\eta=\frac{(1+V)^2a^2f^2}{A^2}\,,\label{sol-eta}\\
&p_r=\frac{mA}{\sqrt{ab}f}\Big(1-af\frac{(1+V)^2}{A^2}\Big)^{1/2}=\frac{mA}{ab}\frac{\dot{r}}{f^2}\label{sol-pr}\,,
\end{align}
\end{subequations}
where $A>0$ is the integral constant.
To have a solution, the increase of the potential $V$ should be slow enough such that $af(1+V)^2<A^2 $.

According to Suskind proposal \cite{Susskind:2018tei}, particle falling in the black hole corresponds to
 chaotic system's evolution. Explicitly the growth of the Rindler momentum of the falling particle is dual to the growth of the operator size of the
 chaotic system. Thus, we shall focus on the
Rindler momentum $p_\rho$ which relates to the radial momentum (near horizon) as,
\begin{align}
\label{prhov0}
p_\rho\sim\sqrt{f}\,p_r\sim \frac{\dot{r}}{ab f^{3/2}}\,.
\end{align}

Before analyzing specific case, we first calculate the behavior of Rindler momentum  near the horizon.
Near the horizon, the functions $f,a,b$ are expressed approximately as
\begin{align}
f(r)\approx f'(r_+)(r-r_+)=\dfrac{4\pi T}{\sqrt{a_+b_+}}(r_+-r)\,,\quad a(r)\approx a(r_+)\equiv a_+\,,\quad  b(r)\approx b(r_+)\equiv b_+\,.
\end{align}
In the above expressions, $T$ is the Hawking temperature, which is
\fa
T=-\sqrt{a_+b_+}\, f'(r_+)/(4\pi)\,.
\ffa
Then, the particle motion \eqref{sol-r} and the Rindle momentum  \eqref{prhov0} behave as
\begin{subequations}
\begin{align}
&\dot{r}\approx4\pi T(r_+-r)\Big(1-4\pi T(r_+-r)\sqrt{\frac{a_+}{b_+}}\frac{(1+V)^2}{A^2}\Big)^{1/2}\,,\label{sol-r2}\\
&p_\rho\sim  (r_+-r)^{-3/2}\dot{r}\label{sol-m2}\,,
\end{align}
\end{subequations}
where we neglect some coefficients in $p_\rho$ expression which is irrelevant to its growth rate.
We are particularly interested in two cases: $V(r)$ is regular and singular at $r_+$.

When $V(r)$ vanishes or is regular at the horizon , i.e., $V(r_+)<\infty$, as $r\rightarrow r_+$,
 the particle motion is dominated by
\begin{align}
\dot{r}\approx4\pi T(r_+-r)\;\Rightarrow\; r_+-r=e^{-4\pi T t}\,.
\end{align}
Then, we obtain the growth of the Rindler momentum \eqref{sol-m2}:
\begin{align}
\underbrace{p_\rho\sim e^{2\pi T t}}_{\text{Rindler momentum growth}}\xrightarrow{~~\text{dual}~~} \underbrace{e^{\lambda t}}_{\text{chaotic operator size growth}}
\end{align}
According to the duality, we find the growth of the operator size saturate the Lyapunov exponent $\lambda=\lambda_L=2\pi/\beta$, $\beta=1/T$,
which restores the Susskind's results.

When $V(r)$ is singular, however, the Rindler momentum  growth may be suppressed,
which corresponds to a suppression of the chaos. Consider a simple kind of singular potential
$V(r)$ behaving as $V(r)=1/\sqrt{r_+-r}$ near horizon, then the particle motion \eqref{sol-r2} becomes
\begin{align}
\dot{r}\approx4\pi T(r_+-r)\gamma\,,\qquad \gamma\equiv  \Big(1-\frac{4\pi T}{A^2}\sqrt{\frac{a_+}{b_+}}\Big)^{1/2}<1\,.
\end{align}
The solution of the above EOM is $r_+-r=e^{-\gamma 4\pi T}$, which leads to the $p_\rho$ growing as
\begin{align}
\underbrace{p_\rho\sim e^{\gamma 2\pi T t}}_{\text{Rindler momentum growth}}\xrightarrow{~~\text{dual}~~} \underbrace{e^{\lambda t}}_{\text{chaotic operator size growth}}
\end{align}
It means the growth of the operator size $\lambda=\gamma \lambda_L<\lambda_L$, namely, the chaos is suppressed.

From the above analysis of the particle motion near the horizon and the chaos,
we find that a singular potential may capture some mechanics of the chaos suppresion based on the duality between particle
motion and operator size. In the following sections, we
further confirm the above picture in explicit gravity background.

\section{Particle falling in hyperscaling violating black brane}

\subsection{Particle free falling}

In this section, we study the particle falling in $4$ dimensional neutral HV black brane geometry,
which takes the form \cite{Dong:2012se,Alishahiha:2012qu}
\begin{subequations}
\label{HV-geo}
\begin{align}
&ds^2=r^\theta\Big(\frac{-f(r)}{r^{2z}}dt^2+\frac{dr^2}{f(r)r^2}+\frac{dx^2+dy^2}{r^2}\Big)\,,
\label{ds}
\\
&
\begin{cases} f(r)=1-(r/r_+)^{2+z-\theta}\,,&\theta\neq2 \\
f(r)=r^{2z-2}(1-(r/r_+)^{2-z})\,, &\theta=2\\
\end{cases}\,.
\label{fr}
\end{align}
\end{subequations}
The horizon of the black brane locates at $r_+$ and the boundary is at $r\rightarrow0$.
$z$ and $\theta$ are the Lifshitz dynamical exponent and hyperscaling violating exponent, respectively.
The constraints from the null energy condition give
\begin{subequations}
\label{NEC}
\begin{align}
&(2-\theta)(2(z-1)-\theta)\geq 0\,,
\label{NEC1}
\\
&(z-1)(2+z-\theta)\geq 0\,.
\label{NEC2}
\end{align}
\end{subequations}
The Hawking temperature is given by
\begin{align}
T=-\frac{r_+^{2-2z}}{4\pi}f'(z_+)=\begin{cases}\dfrac{2+z-\theta}{4\pi r_+^z}\,,&\theta\neq2\\
\dfrac{2-z}{4\pi r_+}\,, &\theta=2\\
\end{cases}\,.
\end{align}

We firstly study the particle free falling in the HV black brane by setting the potential $V(r)=0$.
The particle motion \eqref{sol-r} and its Rindler momentum \eqref{prhov0} now reduce to
\begin{subequations}
\begin{align}
&\dot{r}=\frac{f}{A}r^{\theta/2-2z+1}(A^2r^{2z-\theta}-f)^{1/2}\,,\label{drdt}\\
&p_\rho\sim\frac{r^{2z-2}}{f^{3/2}}\dot{r}\,.\label{prho}
\end{align}
\end{subequations}

For the special case $\theta=2$ and $z<2$, Eq. \eqref{drdt} can be solved analytically, and then we can obtain the analytical expression of $p_\rho$, which reads
\begin{align}
p_\rho(t)=
\sinh\left(\frac{2-z}{2} t \right) \,,
\end{align}
where we have taken $r_+=A=1$ for simple.
It is easy to find that at the late times the Rindler momentum exhibits Lyapunov exponential growth $\lambda=2\pi/\beta$.

\begin{figure}[ht]
\center{
\includegraphics[scale=0.7]{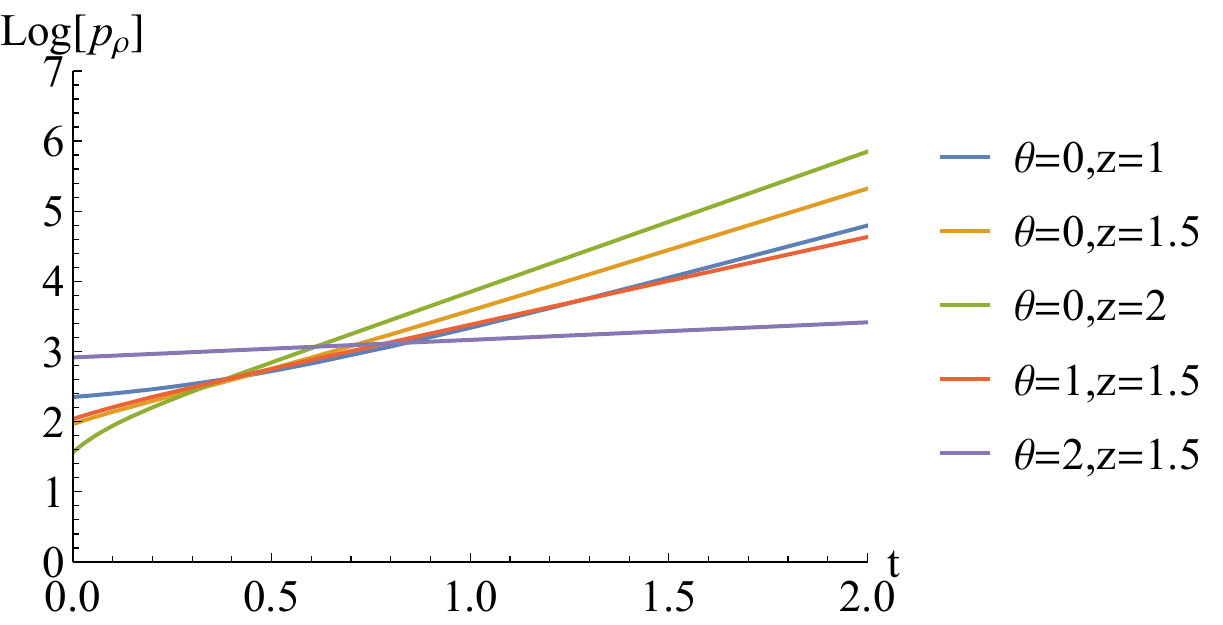}\ \hspace{0.8cm}\ \\
\caption{\label{prvst} The time dependence of Rindler momentum $p_{\rho}(t)$ for different hyperscaling and Lifshitz parameters $\theta$ and $z$.
We set $m=1$, $A=10$ and $r(0)=0.5$ to solve Eq. \eqref{drdt} of the particle.}}
\end{figure}

For general $\theta$ and $z$, it is hard to obtain the analytic solutions of Eq. \eqref{drdt} and we resort to the numerical method instead.
FIG.\ref{prvst} shows the momentum time dependence $p_{\rho}(t)$ for different hyperscaling and Lifshitz parameters $\theta$ and $z$.

The exponential growth of the Rindler momentum at late time is obvious. The growth rate  can
be worked out by locating the slope of $\ln p_{\rho}(t)$.
The results are exhibited in the third line in TABLE \ref{table1}.
Comparing with the Lyapunov exponent $\lambda_L$ (the second line in TABLE \ref{table1}), we conclude that particle falls in HV black hole fastest when
the potential vanishes. It
corresponds to the bound  of the chaotic operator size.

\begin{widetext}
\begin{table}[ht]
\begin{center}
\begin{tabular}{|c|c|c|c|c|c|}
         \hline
~$ $~ &~$\theta=0,z=1$&~$\theta=0,z=1.5$~&~$\theta=0,z=2$&~$\theta=1,z=1.5$&~$\theta=2,z=1.5$
          \\
        \hline
~$\lambda_L=2\pi T$~ & ~$1.5$& ~$1.75$~&~$2$& ~$1.25$&~$0.25$
          \\
        \hline
~$\lambda$(regular $V(r)$)~ & ~$1.49$& ~$1.75$~&~$2.00$& ~$1.25$&~$0.25$
          \\
        \hline
~$\lambda$ (singular $V(r)$)~ & ~$1.25$&~$1.45$~&~$1.43$& ~$1.13$&~$0.24$
          \\
        \hline
\end{tabular}
\caption{\label{table1} The chaos bound $\lambda_L$, the Lyapunov exponent $\lambda$ of the particle free falling and that with repulsive potential for different hyperscaling violation and Lifshitz exponent.}
\end{center}
\end{table}
\end{widetext}

In summary, for the free falling particle, the momentum grows as $p_{\rho}\sim e^{2\pi T t}$,
which is independent of the hyperscaling and Lifshitz parameters $\theta$ and $z$.
It coincides to the analysis in Sec.\ref{basis} where the potential vanishes near horizon.

\subsection{Particle falling with repulsive potential}

Now, we turn on the potential $V$. Then Eq.\eqref{sol} and \eqref{prhov0} in HV background become
\begin{subequations}
\label{solHVp}
\begin{align}
&\dot{r}=r^{1-z}f\Big(1-\frac{r^{\theta-2z}f(1+V)^2}{A^2}\Big)^{1/2}\,,\label{solHVp-drdt}\\
&p_\rho\sim mA\frac{r^{z-1}}{\sqrt{f}}\Big(1-\frac{r^{\theta-2z}f(1+V)^2}{A^2}\Big)^{1/2}\,.
\end{align}
\end{subequations}
Inspired by the conclusion in Sec.\ref{basis} that when the potential $V(r)$ is singular, the
growth of the Rindler momentum $p_\rho(t)$ may be suppressed which is dual to the chaos suppression.
To check this point, we choose the form of the potential as
\begin{align}
\label{rpotential}
V(r)=\sqrt{\frac{1}{f}}-1\,.
\end{align}
which vanishes at the boundary and  is singular at the horizon.
Then, \eqref{solHVp} becomes
\begin{subequations}
\label{solHVp1}
\begin{align}
&\dot{r}=r^{1-z}f\Big(1-\frac{r^{\theta-2z}}{A^2}\Big)^{1/2}\,,\label{solHVp-drdt1}\\
&p_\rho\sim \frac{r^{z-1}}{\sqrt{f}}\Big(1-\frac{r^{\theta-2z}}{A^2}\Big)^{1/2}\,.
\end{align}
\end{subequations}

\begin{figure}[ht]
\center{
\includegraphics[scale=0.7]{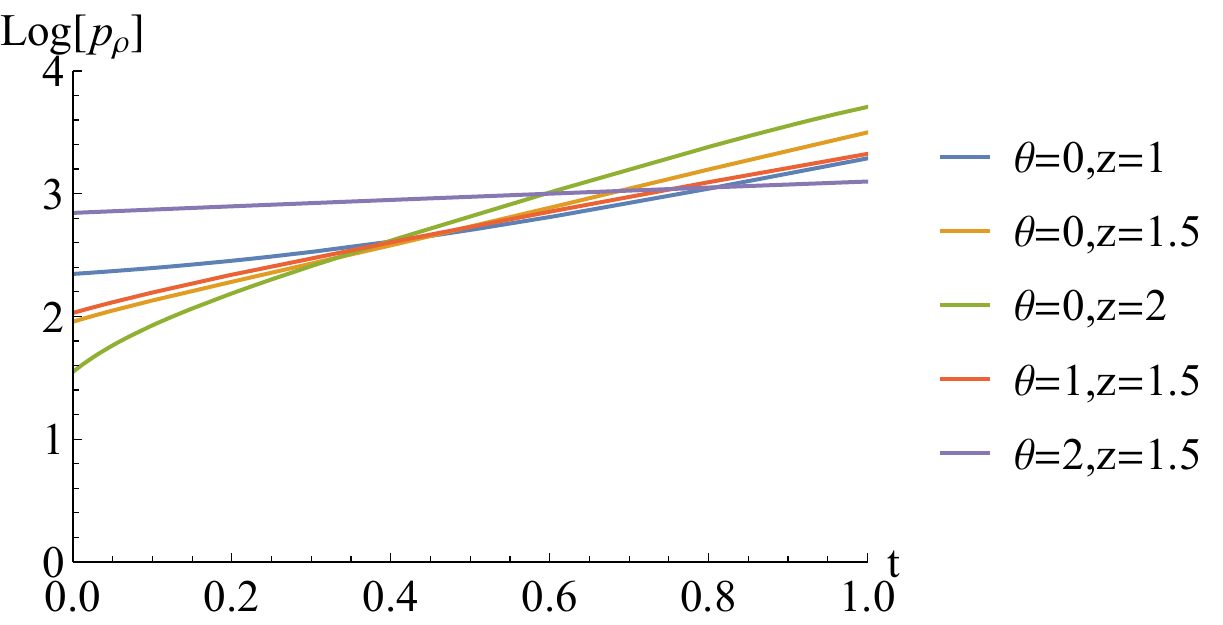}\ \hspace{0.8cm}\ \\
\caption{\label{prvst-p} The time dependence of Rindler momentum  $p_{\rho}(t)$ for different hyperscaling and Lifshitz parameters $\theta$ and $z$.
We set $m=1$, $A=10$ and $r(0)=0.5$ to solve Eq. \eqref{solHVp-drdt1}.}}
\end{figure}

Solving Eq.\eqref{solHVp-drdt1} numerically,
we show the growth of  $p_\rho$ for different $\theta$ and $z$ in FIG\ref{prvst-p}. Although
$p_\rho$ still grows exponentially, the slope of $\ln p_\rho$ shown in the fourth line of TABLE \ref{table1} manifest
a suppression of the growth rate, lower than the  bound $\lambda_L=2\pi T$. Thus, a singular potential like
~\eqref{rpotential} may correspond to the chaos suppression which is below the chaos bound.

\section{Particle falling in AdS soliton background}

In this subsection, we turn to consider the free particle falling in $5$ dimensional AdS soliton geometry \cite{Witten:1998zw,Horowitz:1998ha,Nishioka:2009zj,Horowitz:2010jq}, the metric is given by
\begin{subequations}
\label{Soliton-geo}
\begin{align}
&ds^2=\frac{1}{r^2}\Big(\frac{dr^2}{f(r)}-dt^2+dx^2+dy^2+f(r)d\chi^2\Big)\,,
\\
&f(r)=1-\frac{r^4}{r_+^4}\,,
\end{align}
\end{subequations}
where $r_+$ is the tip of this geometry. To avoid a conical singularity at the tip
we must impose a period of $\chi\sim\chi+\frac{\pi}{r_+}$.
We can make a double Wick rotation of the AdS Schwarzschild black brane
to obtain the geometry \eqref{Soliton-geo}.

Taking the ansatz $r=r(\tau), x=y=\chi=const.$ with the static gauge $\tau=t$ and using  \eqref{eom} with zero potential,
we can obtain the the free particle motion in AdS soliton
\begin{align}
\label{drdt-soliton}
\frac{dr}{dt}=\Big(f-\frac{f}{A^2r^2}\Big)^{1/2}\,.
\end{align}
The Rindler momentum $p_\rho(t)$  \eqref{prhov0} reduces to
\begin{align}
p_\rho=A m \sqrt{1-\frac{1}{A^2 r^2 }}\,.
\end{align}

\begin{figure}[ht]
\center{
\includegraphics[scale=0.7]{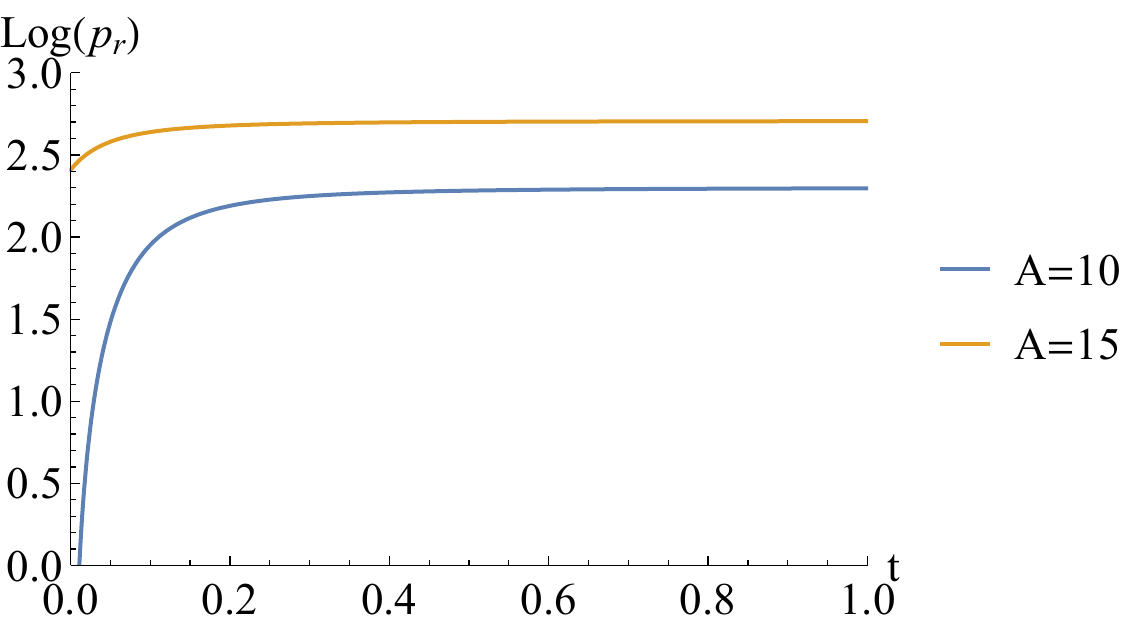}\ \hspace{0.8cm}\ \\
\caption{\label{prvst-soliton} The radial momentum time dependence $p_{\rho}(t)$ in the AdS soliton background.
We set $m=1$ and $r(0)=0.1$ to solve the EOM \eqref{drdt-soliton}.}}
\end{figure}

We numerically solve Eq.\eqref{drdt-soliton} and obtain the momentum time dependence $p_\rho(t)$ (FIG.\ref{prvst-soliton}).
Since the horizon is absent in the AdS soliton background, the chaos is suppressed.
Therefore, the growth of the particle's momentum in AdS soliton geometry
is different from that in the black holes background.
More detailed explorations deserve further studying.

\section{Conclusions and discussions}\label{sec-con}

In this note, we study particle falling in the black hole with an external potential.
By a general analysis near the horizon, we show that the growth rate of the particle's Rindler momentum of the particle falling with a regular potential is as the same as that of the particle free falling, which saturates the chaos bound.
However, when the potential is singular, the growth rate is suppressed such that it is below the Lyapunov exponent.
It means that the chaos suppression may be captured
by such external singular potential in the gravity side.

Further, we explicitly study the momentum growth of a particle when it falls in HV black brane and the soliton background.
For HV case, when the particle free falls toward the black brane, the momentum always grows exponentially, independent of the Lifshitz and HV exponents.
The exponent growth rate $2\pi T$ can be explained by the correspondence between the light speed in gravity and the chaos bound in quantum complexity.

However, after a repulsive potential is introduced,
the exponent of the growth of the momentum may not saturate the bound,
since the the falling particle is slow down and can not achieve the light speed.
The similar situation  happens when the free particle falls in the soliton background. Without a horizon, the light speed is also hard to achieve and then no exponent growth appears.
In a dual manner, we can say that the chaos of the quantum system is suppressed.

Our results explicitly illuminate that the velocity bound in the gravity corresponds to the chaos bound in the quantum system.
To have an well understanding on this point, it is help to specify the line element of the particle trajectory seen by a comoving observer.
Namely, we choose parameter $\lambda$ that $dt/d\lambda=\sqrt{-g^{tt}}$ and express the line element  as
\begin{align}
l^2(\lambda)=g_{\alpha\beta}\frac{d X^\alpha}{d\lambda} \frac{d X^\beta}{d\lambda} =-1+g_{rr}(\frac{dr}{d\lambda})^2+\cdots=\frac{(1+V)^2g_{tt}}{A^2}\,.
\end{align}
When $l^2(\lambda)\rightarrow0$, we say that the particle travels with a light speed to the comoving observer.
Then, for a free particle falling near the black hole horizon,
its speed approaches the light speed as $g_{tt}\rightarrow0$. When a repulsive potential  $V\sim 1/\sqrt{-g_{tt}}$ (as in \eqref{rpotential} ) is introduced,
or the black hole horizon is absent, which results in $g_{tt}\neq0$, $l^2(\lambda)$ is finite and so the growth rate can not reach the chaos bound any more.

Summing up, a repulsive potential or the soliton geometry can provide a platform for us to holographically study the realistic quantum chaos system.
But until now, we cannot explicitly give which specific quantum system being dual to our gravity model.
It is also the difficulties that holography need to work out.
But at least, our study can provide us clues and insights into the realistic quantum chaos system.
Especially, we hope to give some universal properties of the holographic dual theory and explore the basic principle behind these phenomena observed in our gravity side.
In future, we will further pursuit these problems.
The first step is to give an holographic effective field theory, for which the Lyapunov exponent of the dual theory is suppressed.

\begin{acknowledgments}

This work is supported by the Natural Science
Foundation of China under Grant Nos. 11775036, 11747038, 11847313, 11905182.
J. P. Wu is also supported by Top Talent Support Program from Yangzhou University.

\end{acknowledgments}

\end{document}